\documentclass[final,3p,times]{elsarticle}
\usepackage{graphicx}
\usepackage{amssymb}
\usepackage{amsmath}
\usepackage{hyperref}
\usepackage{color}
\usepackage{dsserif}
\usepackage[normalem]{ulem}
\journal{Physics Letters A}
\DeclareUnicodeCharacter{2212}{-}

\usepackage{hyperref}
\usepackage[T1]{fontenc} 

\journal{Computer Physics Communications}

\begin{document}

\begin{frontmatter}

\title{OpenMP Fortran programs for solving the time-dependent dipolar Gross-Pitaevskii equation }

\author[car]{Luis E. Young-S.}
\ead{lyoung@unicartagena.edu.co}

\author[bdu]{Paulsamy Muruganandam}
\ead{anand@bdu.ac.in}

\author[scl]{Antun Bala\v{z}}
\ead{antun@ipb.ac.rs}

\author[ift]{Sadhan K. Adhikari\corref{author}}
\ead{sk.adhikari@unesp.br}
\cortext[author]{Corresponding author.}

\address[car] {Grupo de Modelado Computacional, Facultad de Ciencias Exactas
y Naturales, Universidad de Cartagena,\\ 130014 Cartagena, Bolivar,
Colombia}

\address[bdu]{Department of Physics, Bharathidasan University, Palkalaiperur Campus, Tiruchirappalli 620024, Tamilnadu, India}

\address[scl]{Institute of Physics Belgrade, University of Belgrade, Pregrevica 118, 11080 Belgrade, Serbia}
 
\address[ift]{Instituto de F\'{\i}sica Te\'{o}rica, UNESP -- Universidade Estadual Paulista, 01.140-70 S\~{a}o Paulo, S\~{a}o Paulo, Brazil}

\begin{abstract}
In this paper we present Open Multi-Processing (OpenMP) Fortran 90/95 versions of previously published numerical programs for solving the dipolar Gross-Pitaevskii (GP) equation including the contact interaction in one, two and three spatial dimensions.  The atoms are considered to be polarized along the $z$ axis and we consider  different cases, e.g., stationary and non-stationary solutions of the GP equation for a dipolar Bose-Einstein condensate (BEC) in one dimension  (along $x$ and $z$  axes), two dimensions (in $x$-$y$ and $x$-$z$ planes), and three dimensions. The algorithm used is the split-step semi-implicit Crank-Nicolson scheme for imaginary- and real-time propagation to obtain stationary states and BEC dynamics, respectively,  as in the previous version [R. Kishor Kumar et al., Comput. Phys. Commun. 195, 117 (2015)]. These OpenMP versions have significantly reduced execution time in multicore processors.    
\end{abstract}


\end{frontmatter}

\begin{small}
\noindent
{\bf New version program summary}

\noindent\vspace*{-2mm}\\
{\em Program title:} DBEC-GP-OMP, a program package containing programs imag3d-th.f90, real3d-th.f90, 
imag2dXY-th.f90,  real2dXY-th.f90, imag2dXZ-th.f90, real2dXZ-th.f90, imag1dX-th.f90, real1dX-th.f90,
imag1dZ-th.f90, real1dZ-th.f90,  with fftw3.f03 and fftw3.mod.

\noindent\vspace*{-2mm}\\
{\em CPC Library link to program files:} \url{}

\noindent\vspace*{-2mm}\\
{\em Program obtainable from:} CPC Program Library, Queens University, Belfast, N. Ireland.

\noindent\vspace*{-2mm}\\
{\em Licensing provisions:} Apache License 2.0

\noindent\vspace*{-2mm}\\
{\em Programming language:} Open Multi-Processing (OpenMP)  Fortran 90/95. The program is tested with the GNU, Intel,  and Oracle (former Sun) compilers.


\noindent\vspace*{-2mm}\\
{\em Journal Reference of previous version}: {Comput. Phys. Commun. 195 (2015) 117.} \url{https://doi.org/10.1016/j.cpc.2015.03.024}

\noindent\vspace*{-2mm}\\
{\em Does the new version supersede the previous version?}:  Yes 

\noindent\vspace*{-2mm}\\
{\em Nature of problem:}
The present Open OpenMP Fortran 90/95 programs solve the time-dependent nonlinear partial differential Gross-Pitaevskii (GP) equation for a trapped dipolar Bose-Einstein condensate (BEC) in one (1D), two (2D), and three  (3D) spatial dimensions.

\noindent\vspace*{-2mm}\\
{\em Solution method:}
We employ the split-step Crank-Nicolson scheme to discretize the time-dependent GP equation in space and time. The discretized equation is then solved by imaginary- or real-time propagation, employing adequately small space and time steps, to yield the solution of stationary and non-stationary problems, respectively.

\noindent\vspace*{-2mm}\\
{\em Reason for new version:}
 Previously  published Fortran programs \cite{bec2009} for solving the GP
equation for a BEC
have become useful tools. These programs have been translated to the C programming language  \cite{bec2012} and later extended to the more complex scenario
of dipolar atoms \cite{dbec2015}, spinor condensates \cite{spinor}, and rotating consensates \cite{vor-lat}. Now virtually all computers have multi-core processors and some have motherboards with more than one
physical computer processing unit (CPU), which may increase the number of available CPU cores on a single computer to several
tens. The Fortran \cite{xxx} and  C \cite{bec2016ompmpinondipC} programs for a nondipolar BEC 
have been adopted to be very fast on such multi-core modern computers.  The C programs for a dipolar BEC have been adopted to multicore processors to yield OpenMP, OpenMP/MPI, and CUDA/MPI versions \cite{dbec2016alldipC,dbec2016cudadipC}.

The available  Fortran 90/95 and C programs for the solution of the GP equation for a dipolar BEC  
in 1D, 2D, and 3D  are enjoying widespread use \cite{bec2009,bec2012,dbec2015}. Later,  OpenMP/MPI, and CUDA/MPI  \cite{dbec2016alldipC,dbec2016cudadipC} versions of 
these C programs were published, which highly reduce the CPU times in multicore processors.  Although, Fortran 90/95 is a popular programming language, the previous   Fortran 90/95  programs for  a dipolar BEC \cite{dbec2015}, lacking an efficient parallelization,  suffer from a drawback of requiring large CPU time.
In this paper we present the OpenMP version of the Fortran 90/95 programs  for the solution of the GP equation for a dipolar BEC  
in 1D, 2D, and 3D, 
ideal for an efficient execution in multicore processors, which highly reduce the CPU time.

\noindent\vspace*{-2mm}\\
{\em Summary of revisions:}

The program package DBEC-GP-OMP contains 3D programs  imag3d-th.f90, real3d-th.f90, 2D programs  
imag2dXY-th.f90,  real2dXY-th.f90, imag2dXZ-th.f90, real2dXZ-th.f90,  and 1D programs imag1dX-th.f90, real1dX-th.f90, 
  imag1dZ-th.f90, real1dZ-th.f90, in the directory src, as well as the files makefile,  readme.txt, and
   readme-fftw.txt.  The 3D and 2D programs are OpenMP versions leading to a significant reduction in execution time in a multicore processor. However, in 1D the supplied programs are autoparallel versions as 
   there is no further reduction in the execution time in the  OpenMP versions. 
    The makefile allows an automated compilation of the programs using different supported compilers (GNU, Intel, Sun Oracle) by a simple make command, as in Ref.~\cite{dbec2015}. {However, the 1D and 2D dipolar programs in previous publications \cite{dbec2015,dbec2016alldipC,dbec2016cudadipC} erroneously used the 3D nonlinearities, in place of corresponding 1D and 2D nonlinearities,  in numerical calculations. The corresponding arXiv archive versions do not have these errors.} The file readme.txt contains instructions on how to compile and run the programs.   The file readme-fftw.txt provides instructions for installing the fast Fourier transformation (FFT) routine in Linux/Unix operating systems unless this routine is preinstalled. All input parameters are placed in the beginning of each program in MODULEs GPE$\_$DATA and COMM$\_$DATA. 
The directory output contains examples of matching outputs of imaginary- and real-time propagation programs in sub-directories with a generic name, e.g.  imag3d-th, real2dXY-th, etc. The reader is advised to consult Ref. \cite{dbec2015} for details.  The imaginary-time programs implement the calculation starting from an initial analytic Gaussian wave function using a non-zero value of the parameter NSTP, whereas the real-time programs use the converged solution of the imaginary-time programs as the initial state of calculation employing NSTP = 0.
 The FFT algorithm works faster and more efficiently when we take the number of space grid points $-$ N in 1D and NX, NY, NZ in 2D and 3D $-$ in powers of 2, e.g. $2^n$ with $n$ an integer, and should be so chosen for reducing the execution time. 
The operational scheme for running the codes is identical  to that in Ref. \cite{dbec2015}.

 \end{small}

\section*{Acknowledgments}
\noindent
L.E.Y.-S.  acknowledges the financial support
by the Vicerrectoria de Investigaciones, Universidad de Cartagena, Colombia  through Project No. 019-2021.
 P.M. acknowledges the financial support    
 by the Council of Scientific and Industrial Research (CSIR), India under Grant No. 03(1422)/18/EMR-II, and Science and Engineering Research Board (SERB), India under Grant No. CRG/2019/004059. A.B. acknowledges funding provided by the Institute of Physics Belgrade, through a grant by the Ministry of Education, Science, and Technological Development of the Republic of Serbia. 
S.K.A. acknowledges support by the CNPq, Brazil through  grant 301324/2019-0.

\end{document}